\documentclass[printer]{aa}  
\usepackage{caption}
\usepackage{subcaption}
\usepackage{graphicx}
\usepackage{hyperref}
\usepackage{amsmath}
\hypersetup{
  colorlinks   = true, %Colours links instead of ugly boxes
  urlcolor     = red, %Colour for external hyperlinks
  linkcolor    = blue, %Colour of internal links
  citecolor   = blue %Colour of citations, could be ``red''
}
\usepackage{txfonts}
\newcommand{\ts}{\textsuperscript}
\begin{document}

   \title{On the short term stability and tilting motion \\ of a well-observed low-latitude solar coronal hole}

   \author{Stephan~G.~Heinemann
          \inst{1}
          \and
          Stefan~J.~Hofmeister\inst{2}
          \and
          James~A.~Turtle\inst{3}
          \and
          Jens~Pomoell\inst{1}
          \and
          Eleanna~Asvestari\inst{1} 
          \and
          Alphonse~C.~Sterling\inst{4} 
          \and
          Andrea~Diercke\inst{5} 
          \and
          Cooper~Downs\inst{3} 
        }

   \institute{Department of Physics, University of Helsinki, P.O. Box 64, 00014, Helsinki, Finland\\
    \email{stephan.heinemann@hmail.at, stephan.heinemann@helsinki.fi}
         \and
            Leibniz-Institut f\"ur Astrophysik Potsdam (AIP), 
             An der Sternwarte 16, 
             14482 Potsdam, Germany
         \and
             Predictive Science Inc., 9990 Mesa Rim Road, Suite 170, San Diego, CA 92121, USA
         \and
             NASA/Marshall Space Flight Center, Huntsville, AL 35812, USA
         \and
             Leibniz-Institut f\"ur Sonnenphysik (KIS), 
            Sch\"oneckstr. 6,
            79104 Freiburg, Germany
             }

   \date{Accepted September 15, 2023}
% \abstract{}{}{}{}{} 
% 5 {} token are mandatory
 
  \abstract
   {The understanding of the solar magnetic coronal structure is tightly linked to the shape of open field regions, specifically coronal holes. A dynamically evolving coronal hole coincides with the local restructuring of open to closed magnetic field, which leads to changes in the interplanetary solar wind structure.}
   {By investigating the dynamic evolution of a fast-tilting coronal hole, we strive to uncover clues about what processes may drive its morphological changes, which are clearly visible in extreme ultraviolet (EUV) filtergrams.}
   {Using combined 193\AA\ and 195\AA\ EUV observations by the Atmospheric Imaging Assembly on-board the Solar Dynamics Observatory and the Extreme Ultraviolet Imager on-board the Solar Terrestrial Relations Observatory - Ahead, in conjunction with line-of-sight magnetograms taken by the Helioseismic and Magnetic Imager, also on-board the Solar Dynamics Observatory, we track and analyze a coronal hole over 12 days to derive changes in morphology, area and magnetic field. We complement this analysis by potential field source surface modeling to compute the open field structure of the coronal hole. }
  % results heading
   {We find that the coronal hole exhibits an apparent tilting motion over time that cannot solely be explained by solar differential rotation. It tilts at a mean rate of $\sim 3.2^{\circ}\mathrm{day}^{-1}$ that accelerates up to $\sim 5.4^{\circ}\mathrm{day}^{-1}$. At the beginning of May, the area of the coronal hole decreases by more than a factor of three over four days (from $\sim 13 \times 10^{9}$\,km$^{2}$ to $\sim 4 \times 10^{9}$\,km$^{2}$), but its open flux remains constant ($\sim 2\times 10^{20}$\,Mx). Further, the observed evolution is not reproduced by modeling that assumes the coronal magnetic field to be potential.}
   {In this study, we present a solar coronal hole that tilts at a rate that has yet to be reported in literature. The rate exceeds the effect of the coronal hole being advected by either photospheric or coronal differential rotation. Based on the analysis we find it likely that this is due to morphological changes in the coronal hole boundary caused by ongoing interchange reconnection and the interaction with a newly emerging ephemeral region in its vicinity.}


   \maketitle
   %\tableofcontents
%
%-------------------------------------------------------------------

\section{Introduction}

\begin{figure*}[h!t]
   \centering
   \includegraphics[width=0.9\linewidth]{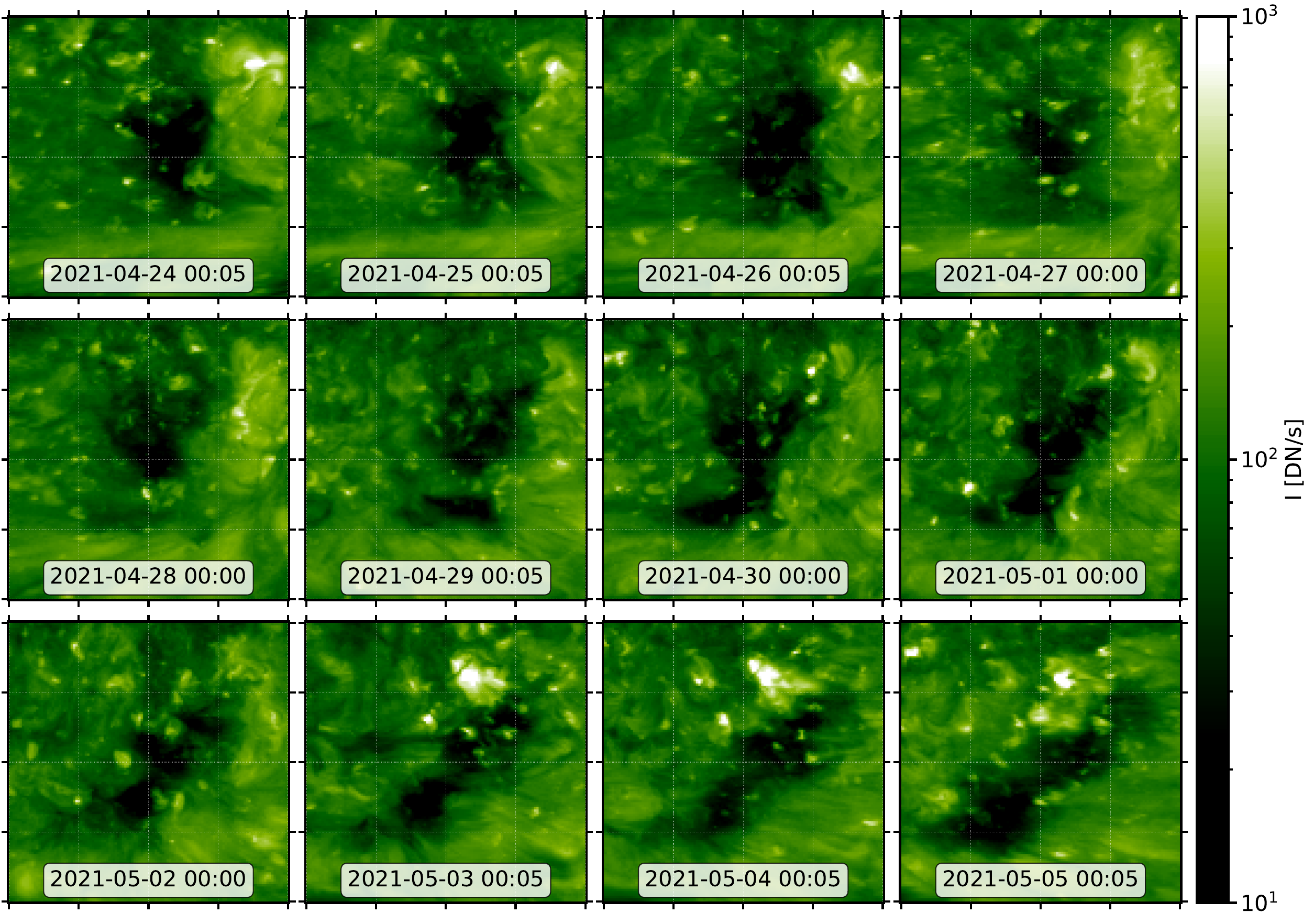}

   \caption{Daily EUV images featuring the coronal hole. The consecutive panels show EUV images that highlight the evolution of the coronal hole between April 24\ts{th} and May 5\ts{th} 2021. The images are prepared by merging STEREO-A/EUVI 195\AA\ and SDO/AIA 193\AA\ filtergrams in the CHMAP pipeline. The coronal hole is shown in the heliographic \textit{Carrington} frame on a $10^{\circ}$ grid (white guidelines) from $-50^{\circ}$  to $-10^{\circ}$ latitude and $120^{\circ}$  to $160^{\circ}$ longitude. }
              \label{fig:overview}%
\end{figure*}

%introduction to open fields
The topology and dynamics of the solar atmosphere are governed by the complex interplay of open and closed magnetic fields. Open fields link the solar surface to interplanetary space, allowing outflowing plasma to be accelerated to supersonic speeds, and are the source of the heliospheric open flux \citep[see review by][]{2019cranmer}. Open field structures lead to the formation of large-scale, low-density, low-temperature regions in the solar corona that can be observed as structures of reduced emission in extreme ultraviolet (EUV) and X-rays (on-disk) and also in white-light (off-limb), --- the so-called coronal holes \citep[see review by ][and references therein]{cranmer2009}.\\

%general introduction to CHs
Due to plasma outflow along open magnetic field lines, coronal holes feature an electron density that can be more than $30\%$ lower than the surrounding quiet (closed) corona \citep[($1~-~2.5)~\times~10^{8}~$cm$^{-3}$;][]{1999warren,2011hahn,2021heinemann_dem} and a temperature around $0.9$~MK \citep[\textit{e.g.,}][]{1999Fludra,2020saqri}. The magnetic field structure of coronal holes consists of a mix of open and closed magnetic fields. The open fields form magnetic funnels \citep{2005tu} that are rooted in photospheric footpoints. These footpoints are unipolar magnetic elements of the coronal hole's dominant polarity. Those magnetic elements contain the majority of the coronal hole's signed flux ($38-72\%$) in only a fraction of the coronal hole's area \citep[$1-8\%$;][]{2017hofmeister,2019hofmeister,2019heinemann_catch}. Most of the area of the coronal hole is covered by small closed loops whose average height is significantly lower than in quiet Sun regions \citep{2004wiegelmann}. The interplay of these open and closed fields within the coronal hole and the surrounding and global magnetic field configuration determine the properties of the resulting high-speed solar wind stream in terms of peak velocity, longitudinal and latitudinal extent, shape of the solar wind profile, and whether the stream is geoeffective \citep[\textit{e.g.,} see][]{Garton2018_hss,Geyer2021,Hofmeister2020,Hofmeister2022,Samara2022}. During solar maximum, well-defined low-latitude coronal holes --- that often, but not exclusively, emerge from the ``ashes'' of decaying active regions \citep{2010Karachik,2013Petrie} --- dominate the solar corona, whereas during solar minimum large-scale polar coronal holes appear as a result of flux transport to the poles and the ensuing strong dipole field \citep[][and references therein]{cranmer2009}. Large fragmented coronal holes with diffuse boundaries (\textit{i.e.,} ``patchy'' coronal holes) also often appear at low latitudes during solar minimum \citep{2020Heinemann_chevo,Samara2022}.\\

%intro to coronal hole evolution
To understand and model the coronal structure in a time-dependent fashion, the evolution of coronal holes must be analyzed. The evolution of the area of long-lived coronal holes typically exhibits a pattern of growing to a maximum before decaying over multiple solar rotations with lifetimes from a few solar rotations up to more than 2 years \citep{2018heinemann_paperI,2020Heinemann_chevo,Hewins2020}. \cite{wang90} and \cite{gosling96} argue that changes in the underlying photospheric magnetic field affect the appearance of coronal holes in the corona. \cite{2020Heinemann_chevo} however, showed that there is no systematic correlation between the evolution of the underlying magnetic field and the area of the coronal hole, suggesting that the primary driver of the coronal hole's evolution is the interaction with the local magnetic field configuration surrounding the coronal hole and the global magnetic field.\\

On small scales, the coronal hole boundary changes dynamically through interchange reconnection, also called ``footpoint switching'', where surrounding closed fields interact with the coronal hole's open fields. This reconfiguration may cause a coronal hole to grow or decay. The rate of this process seems to be largely reliant on the surrounding fields, \textit{i.e.,} abundance of large scale arcades and nearby active regions \citep[e.g., see][]{1984Shelke, 2004wang, 2004Madjarska, 2009madjarska, 2011yang, 2014ma,2018kong}. Studies show that coronal holes evolve continuously (as long as external drivers, such as a rapid global reconfiguration of the coronal magnetic field, are not apparent), which suggests that interchange reconnection processes occur regularly at the coronal hole boundaries \citep{Bohlin1977,1990Kahler}. Interchange reconnection due to emerging flux within the coronal hole may, however, further aid in the decay of coronal holes \citep{2007Zhang}. Recently, interchange reconnection has also been suggested as a driver of the outflowing fast solar wind \citep{bale2023}.\\

%intro to rotation
Coronal holes may also change their shape over the course of their lifetime. Although often considered as rigid structures \citep{1975timothy}, studies have shown that solar differential rotation can have a significant effect on the coronal manifestation of a coronal hole \citep{2016caplan_LBC_IIT,2020Heinemann_chevo}. \cite{Wang1993} proposed that to maintain a force-free state of open field structures that makes up the coronal holes, the rotation needs to be quasi-rigid to avoid field line twisting, but they neglected interchange reconnection.  Although many coronal holes maintain their rough shape for some time, solar differential rotation often results in an apparent counterclockwise (northern hemisphere) or clockwise (southern hemisphere) tilting motion of coronal holes with a north-south extension. Prominent examples are the 1996 Elephant's Trunk coronal hole \citep[as studied by][and others]{1999delzanna} and the well-studied 2012 long-lived coronal hole \citep[described by][]{2018heinemann_paperII,2018heinemann_paperI}.\\

%why .. motivation and what
While coronal holes have been intensively studied, the evolution of the connection between the magnetic field in the photosphere and the resulting observed coronal hole morphology is not yet understood.  Usually, coronal holes are treated as force-free and potential open fields rooted in the photosphere, and yet studies show significant differences in modeled open fields and coronal holes observed in EUV \citep{2019asvestari,2021Linker_ISSI}. In this work we analyzed the evolution of one well-observed on-disk small coronal hole, using EUV images and magnetograms as well as magnetic field extrapolations. We have observed an especially rapid rate of coronal hole tilting motion that has not yet been reported in the literature. This rapid evolution is inconsistent with changes caused by the motion of magnetic flux elements in the photosphere and exceeds effects of differential rotation. Our results provide insight into the connection between the photospheric and coronal field signatures in open-field regions and highlight processes that may be involved in coronal hole evolution and the possible subsequent restructuring of the global field.  In Section~\ref{sec:obs}, we present the coronal hole and its evolution over $12$ days. In Section~\ref{sec:disc}, we explore the results and propose two possible physical mechanisms that may be involved in the apparent accelerated tilting motion. The results are then summarized in Section~\ref{sec:sum}.

\section{Observational Results}
\label{sec:obs} 

\subsection{Data and preparation}
   \begin{table}
      \caption[]{CHMAP extraction parameters.}
         \label{tab:thresh}
     $$ 
         \begin{tabular}{l c c}
            \hline
            \noalign{\smallskip}
            Parameter      &  \cite{2016caplan_LBC_IIT} & This study \\
            &  $\log_{10}(I)$ & $\log_{10}(I)$ \\
            \noalign{\smallskip}
            \hline
            \noalign{\smallskip}
            seeding threshold & 0.95 & 1.105     \\
            outer threshold & 1.35 & 1.35            \\
            \noalign{\smallskip}
            \hline
         \end{tabular}
     $$ 
   \end{table}

In this study, we use data from EUVI \citep[\textit{Extreme UltraViolet Imager;}][]{Wuelser2004_EUVI} and AIA \citep[\textit{Atmospheric Imaging Assembly};][]{2012lemen_AIA}, which are EUV imagers on-board the STEREO-A \citep[Solar TErrestrial RElations Observatories - Ahead;][]{2008kaiser_STEREO} and \citep[\textit{Solar Dynamics Observatory};][]{2012pesnell_SDO} spacecrafts. Since STEREO-A was situated $\sim53$ degrees East of SDO, combining the two datasets increased the observational coverage of the corona during the time range of interest. Of EUVI and AIA, we use the 195\AA\ and 193\AA\ channels because of the reduced emissivity of coronal holes in contrast to the surrounding corona in these narrow bands, leading to a significant contrast. The primary contribution to these filters is the coronal emission line of  eleven times ionized iron (Fe \textsc{xii}) and they are commonly used for coronal hole extraction \citep[\textit{e.g.,}][]{Boucheron2016, 2017hofmeister,2019heinemann_catch}. \\

The EUV images used for the analysis and the coronal hole boundaries were prepared and extracted using the Coronal Hole Mapping and Analysis Pipeline (CHMAP) open-source Python software package\footnote{\url{https://github.com/predsci/CHMAP}}. CHMAP integrates all available instruments, in this case AIA/SDO and EUVI/STEREO-A into a single, synchronic Carrington map with coronal hole detections. Prior to mapping, each image undergoes several preparation steps: 1) GPU-enabled PSF deconvolution, 2) data-driven limb brightening correction, and 3) inter-instrument transformation. The deconvolution step is instrument specific and tends to sharpen features in the disk image and removes a portion of the long-distance scattered light, increasing the dynamic contrast in the images. The limb brightening correction uses the equatorial band of images averaged over a six month period to fit a correction factor as a function of center-to-limb angle and pixel intensity. This correction has the effect of ‘flattening’ image intensity from center to limb while leaving active regions and holes intact. The third correction is a linear transform between instruments. This is again calculated from an equatorial band that is time-averaged over six months and brings images from the two instruments to a similar overall brightness. Once the images from different instruments are brought to a consistent format, coronal hole detections are performed by an iterative, kernel-based scheme with one, lower, intensity threshold for seeding and a second, higher, threshold for boundary definitions \citep{2016caplan_LBC_IIT}. Thus, a coronal hole must have at least one pixel less than the lower threshold. The chosen values are shown in Table~\ref{tab:thresh}. The original detection parameters in \cite{2016caplan_LBC_IIT} were trained on qualitative expert evaluations for the period of 2012-2014.  For this study, the seeding parameter was increased from 0.95 to 1.105 due to increased overall brightness of the corrected images relative to the training period causing some apparent coronal holes to be missed. The algorithm then iteratively grows the hole from these ‘seeds’ until the kernel and/or upper threshold are met at all coronal hole edges. The detection is performed at full disk-image resolution and the images are then interpolated to \textit{Carrington} maps (4096x2048 in longitude and latitude respectively). For each time step, we now have a map for each imaging instrument. These maps are combined into a single synchronic map by a minimum-intensity-merge procedure with consideration for viewing angles.  A map pixel that is on the limb for AIA, but nearly at disk center for STEREO-A, will always get the STEREO-A value. Following the merge, map resolution was reduced by integration to 1080x540, which corresponds to a spatial resolution of $0.33\dot{3}$ degrees. We used EUV images prepared by the CHMAP pipeline from April 23\ts{rd} 2021 (when the coronal hole rotated into STEREO-A's field-of-view) to May 5\ts{th} 2021 (when the coronal hole rotated out of SDO's field-of-view) at a $5$ minute cadence. \\

During the time period when the coronal hole was visible in SDO observations, we used $720$s line-of-sight (LOS) observations from HMI \citep[\textit{Helioseismic and Magnetic Imager};][]{2012schou_HMI} on-board SDO to derive the underlying magnetic flux density and its signed and unsigned flux. The magnetic field data were projected to heliographic coordinates and corrected for solar curvature assuming radial fields according to \cite{2017hofmeister} and \cite{2019heinemann_catch}.\\

\subsection{A tilting coronal hole}
\label{ssec:ch}

Figure~\ref{fig:overview} shows snapshots of the evolution of the coronal hole over 12 days. Even without sophisticated analysis, it is clear that the coronal hole undergoes significant changes during that period of time: The coronal hole displays a strong clockwise ``tilting'' motion with time over the duration of the observations and a decrease in its area. Although the complete dissolution of the coronal hole cannot be fully observed because it rotated out of SDOs field-of-view, we note that (1) the coronal hole is barely visible in EUV after May 5\ts{th} while it was still in the field of view of SDO (even when taking into account projection effects) and (2) no trace of it is found when the corresponding  \textit{Carrington} longitude rotates into the STEREO-A field of view on May 19\ts{th}. In addition to the morphological evolution of the coronal hole itself, we observe changes in the surrounding corona: The decay of an ephemeral region at the north-western edge (around $155^{\circ}$~\textit{Carrington} longitude and $-6^{\circ}$~latitude) that was present at the start of the observations. And the emergence of a second ephemeral region close-by a few days later (around $142^{\circ}$~\textit{Carrington} longitude and $-8^{\circ}$~latitude). The evolution of this region is described in \cite{Diercke2023}. 

\subsection{Morphological evolution}
\label{ssec:chevo}

\begin{figure}
   \centering
   \includegraphics[width=\linewidth]{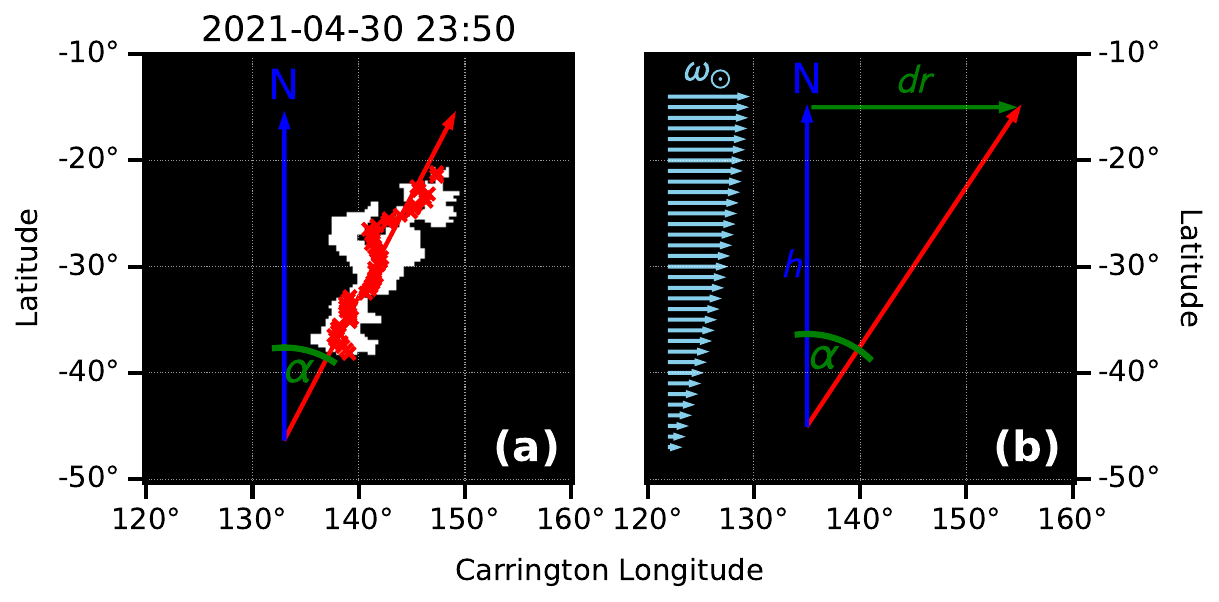}
        
    \caption{Tilt angle calculation. Panel (a) shows an example of the tilt angle calculation for the coronal hole on April 30\ts{th}, 2021 at 23:50. The red crosses mark the latitudinal geometric centers of mass which are then fitted (red line). The angle between the fit and the North-South direction (blue line) is defined as the tilt angle of the coronal hole. In Panel (b) a schematic of the tilt angle calculation solely caused by differential rotation is shown. The light blue arrows highlight the solar differential rotation rates at different latitudes (not to scale).}
                  \label{fig:2}%
\end{figure}

Using the extracted boundaries of the coronal hole, we calculated a tilting rate of the coronal hole by defining the tilt as the angle ($\alpha$) between the major axis of the coronal hole and the north-south axis of the Sun. We determine the major axis by calculating a longitudinal geometric center of mass of the coronal hole extraction at every latitude and then applying a linear fit. An example is shown in Figure~\ref{fig:2}a. From this the tilting rate of the coronal hole is defined as
\begin{equation}\label{eq:alpha_dot}
    \dot{\alpha} = \frac{\partial \alpha}{\partial t}.
\end{equation}
To compare the observed tilting rate of the coronal hole to the differential rotation, we consider an idealized vertical coronal hole (approximated by a line $\equiv$ tilt axis) with a given latitudinal extent ($h$). Due to differential rotation, the axis of such a coronal hole should approximately tilt according to the difference in rotation rate between the lower and upper part. The difference in rotation rate is given by
\begin{equation}\label{eq:delta_omega}
    \Delta\omega_{\odot}=\omega_{\odot}(\lambda_\mathrm{min})-\omega_{\odot}(\lambda_\mathrm{max})
\end{equation}
with $\omega_{\odot}$ denoting the differential rotation rate at a given latitude ($\lambda$). With this relative rotation rate between the lower and upper part of the theoretical coronal hole, we can calculate the relative longitudinal shift $dr$ as function of time (t)
\begin{equation}\label{eq:dr}
    dr(t)=\Delta\omega_{\odot} \times t,
\end{equation}
and subsequently the tilt angle
\begin{equation}\label{eq:alpha}
    \alpha(t) = \tan(\frac{dr}{h}).
\end{equation}
This is schematically shown in Figure~\ref{fig:2}b.\\

To compare our coronal hole tilting rate to an established photospheric rotation rate we use the differential rotation profile given by \cite{Snodgrass1990}, derived from magnetic feature correlation of Mount Wilson magnetograms
\begin{equation}\label{eq:phot}
\omega_{phot} = 14.71 - 2.39 \times \sin^2(\lambda) - 1.78 \times \sin^4(\lambda).
\end{equation}
We approximate the coronal differential rotation by using the profile for coronal hole rotation rates in the southern hemisphere derived by \cite{2017bagashvili}
\begin{equation}\label{eq:coronal}
\omega_{ch} = 14.07 - 0.48 \times \sin^2(\lambda) - 1.71 \times \sin^4(\lambda).
\end{equation}

By assuming a longitudinal extent of $20^{\circ}$ stretching from latitudes $\lambda_\mathrm{min} = -20^{\circ}$ to $\lambda_\mathrm{max} = -40^{\circ}$, we can calculate the tilt angle of a hypothetical coronal hole under the influence of either the photospheric or the coronal differential rotation rate. By substituting the differential rotation profiles (Eq.~\ref{eq:phot} and \ref{eq:coronal}, respectively) into Equation~\ref{eq:delta_omega}, we obtain
\begin{align}
    \Delta\omega_{phot}&=0.99^{\circ}\mathrm{day}^{-1},\\
    \Delta\omega_{ch}&=0.41^{\circ}\mathrm{day}^{-1}. 
\end{align}

\begin{figure*}
   \centering
   \includegraphics[width=\linewidth]{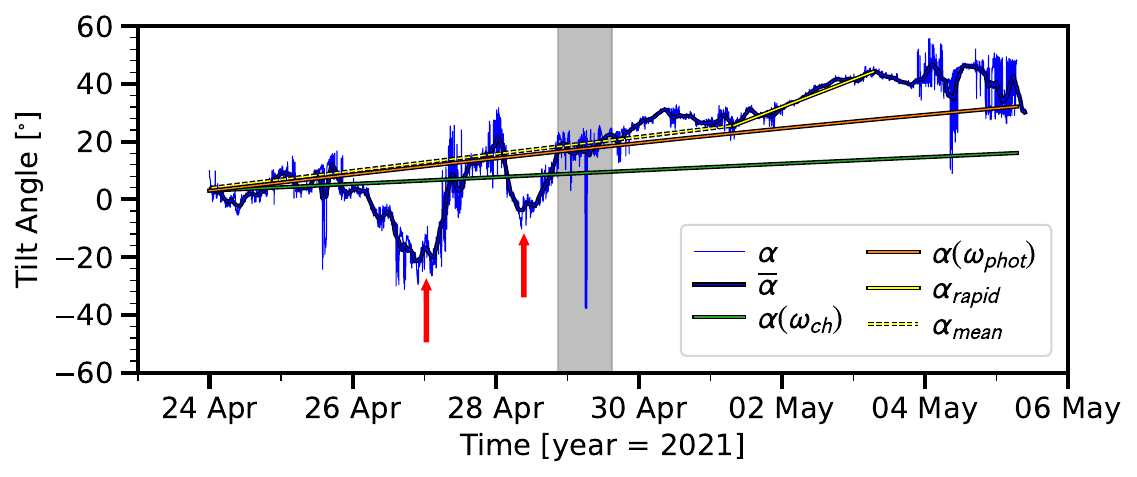}
   \caption{Coronal hole tilt angle ($\alpha$) as functions of time from April 24\ts{th} to May 5\ts{th}. The 3-hour running mean is highlighted in black ($\overline{\alpha}$). The gray shaded area shows the time period of unreliable coronal hole extraction due to the offlimb pointing of SDO (from April 28\ts{th} 21:00 to April 29\ts{th} 15:00). The hypothetical tilt angles caused by advection due to solar differential rotation as function of time are shown as the orange and green lines for photospheric ($\alpha(\omega_{phot})$) and coronal ($\alpha(\omega_{ch})$) differential rotation profiles respectively. The yellow lines highlight the time periods of the mean and rapid tilting rate (as described in Sec.~\ref{ssec:chevo}) and the red arrows mark the short time periods where the tilt angle changed significantly due to localized short-term boundary changes.}
              \label{fig:angle}%
\end{figure*}

Using Equations~\ref{eq:dr} and~\ref{eq:alpha}, we derive the respective tilt angles ($\alpha(\omega_{phot})$ and $\alpha(\omega_{ch})$) as function of time. These are shown in Figure~\ref{fig:angle} as the orange and green lines. The tilting rate can then be derived using Equation~\ref{eq:alpha_dot}, giving hypothetical coronal hole tilting  rates solely caused by advection due to differential rotation of $\dot{\alpha}(\omega_{phot}) = 2.6^{\circ}\mathrm{day}^{-1}$ and $\dot{\alpha}(\omega_{ch}) = 1.2^{\circ}\mathrm{day}^{-1}$.\\

Figure~\ref{fig:angle} shows the coronal hole tilt angle as a function of time between April 24\ts{th} and May 5\ts{th}. On average we find that the coronal hole shows an apparent clockwise tilting motion, as expected for a north-south elongated coronal hole located in the southern hemisphere \citep{2018heinemann_paperI}. Between April 24\ts{th} and May 1\ts{st} the coronal hole tilts at a mean rate of $\sim 3.2^{\circ}\mathrm{day}^{-1}$, although periods of fast counterclockwise followed by immediate fast clockwise tilting are present (marked in Fig.~\ref{fig:angle} by the red arrows). After visual inspection of the coronal hole boundaries (see Fig.~\ref{fig:overview}), we conclude that these short term changes of the tilt angle are caused by localized short term changes in the coronal hole boundary and ambiguity in the determination of the tilt angle rather than the long term evolution of the coronal hole. Interestingly, between May 2\ts{nd} and May 4\ts{th} the tilt angle seems to change at a higher rate of $\sim 5.4^{\circ}\mathrm{day}^{-1}$ (yellow line), but in contrast to the earlier fast changes, it does not seem to be caused by short term boundary displacement but rather due to an apparent tilting motion of the whole coronal hole structure.\\

We find that the mean tilting rate of the coronal hole ($\sim 3.2^{\circ}\mathrm{day}^{-1}$) between April 24\ts{th} and May 1\ts{st} is slightly higher than the photospheric rate ($2.6^{\circ}\mathrm{day}^{-1}$), and significantly larger than the coronal tilting rate ($1.2^{\circ}\mathrm{day}^{-1}$). Between May 2\ts{nd} and May 4\ts{th}, where the coronal hole tilts rapidly, the rate of $\sim 5.4^{\circ}\mathrm{day}^{-1}$ is more than double the photospheric and over four times the coronal rate. \\

\begin{figure}
   \centering
   \includegraphics[width=\linewidth]{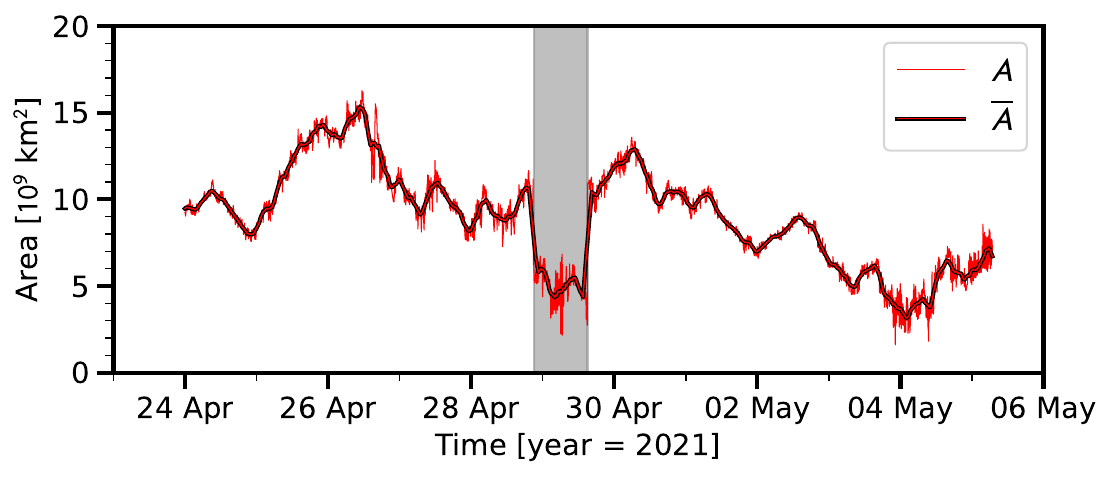}
   \caption{Coronal hole area ($A$) as function of time from April 24\ts{th} to May 5\ts{th}. The 3-hour running mean is highlighted in black ($\overline{A}$). The gray shaded area shows the time period of unreliable coronal hole extraction due to SDOs offlimb pointing (from April 28\ts{th} 21:00 to April 29\ts{th} 15:00).}              \label{fig:area}%
\end{figure}

Figure~\ref{fig:area} shows the coronal hole area as function of time. We observe that the area changes significantly between April 24\ts{th} and April 26\ts{th} by increasing in size by over $50\%$ from $10 \times 10^{9}$ km$^{2}$ to $15 \times 10^{9}$ km$^{2}$ and then decaying back to $10 \times 10^{9}$ km$^{2}$ within a day. Subsequently until around April 30\ts{th} the area seems to remain constant, notwithstanding that the data in the time period between April 28th 21:00 and April 29th 15:00 is unreliable because SDO was pointing off-limb due to a special observation campaign and the STEREO-A data showing strong projection effects. Starting around April 29th 06:00, we observe a steady decline to $\sim 4 \times 10^{9}$ km$^{2}$ until May 4\ts{th}. Finally, we find a slight increase in the area up to $\sim 5 \times 10^{9}$ km$^{2}$ in the hours before the coronal hole rotated out of SDO's field-of-view.\\ 

\subsection{Magnetic evolution}
\label{ssec:chmag}

\begin{figure}
   \centering
   \includegraphics[width=\linewidth]{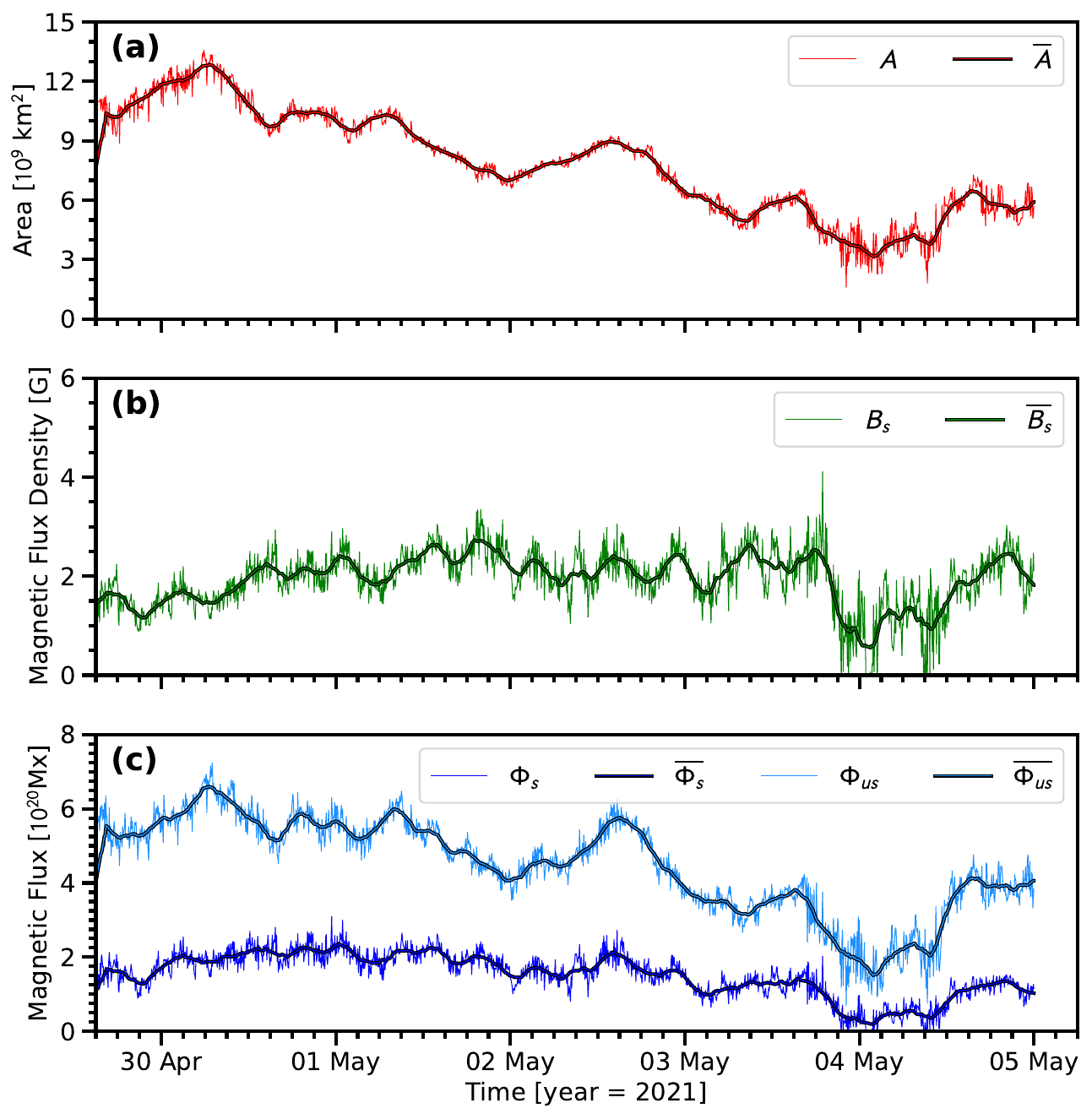}
   \caption{Coronal hole properties as functions of time from April\,29\ts{th}\,15:00 to May\,4\ts{th}\,00:00. The coronal hole area ($A$; red), signed magnetic flux density ($B_{s}$;green), unsigned magnetic flux ($\Phi_{us}$; light blue) and signed magnetic flux ($\Phi_{s}$; blue) are shown from top to bottom with a 3-hour running mean highlighted in black ($\overline{A}$, $\overline{B_{s}}$, $\overline{\Phi_{us}}$ and $\overline{\Phi_{s}}$ respectively).}
              \label{fig:props}%
\end{figure}

Here we present the evolution of the coronal hole's magnetic properties during the time period when the coronal hole was observed by SDO. In Figure~\ref{fig:props} the area, magnetic flux density, and magnetic flux are shown for the time period April 30\ts{th} to May 5\ts{th}. The magnetic field properties were calculated by integrating each magnetogram over the projected coronal hole boundary and correcting for radial fields \citep[for details see][]{2019heinemann_catch}. The area decreases continuously from $\sim 12.8 \times 10^{9}$\,km$^{2}$ on April 29\ts{th} 22:00 to $\sim 3.9 \times 10^{9}$\,km$^{2}$ on May 3\ts{rd} 22:00 (Fig.~\ref{fig:props}, top panel). The area-averaged (signed) magnetic flux density of the underlying photospheric field within the projected coronal hole area (Fig.~\ref{fig:props}, middel panel) increases slowly from $ 1.46 \pm 0.32$ G on April 29\ts{th} to $ 2.26 \pm 0.42$\,G on May 3\ts{rd} (the time-averaged value was found to be $ 2.03 \pm 0.47$\,G). The respective signed (or open) flux remains approximately constant over most of the observed time period between April 29\ts{th} and May 3\ts{rd}  at $(1.82 \pm 0.43) \times 10^{20}$\,Mx.\\

\begin{figure*}
    \centering
    \begin{subfigure}[t]{0.495\textwidth}
        \centering
        \includegraphics[width=\linewidth]{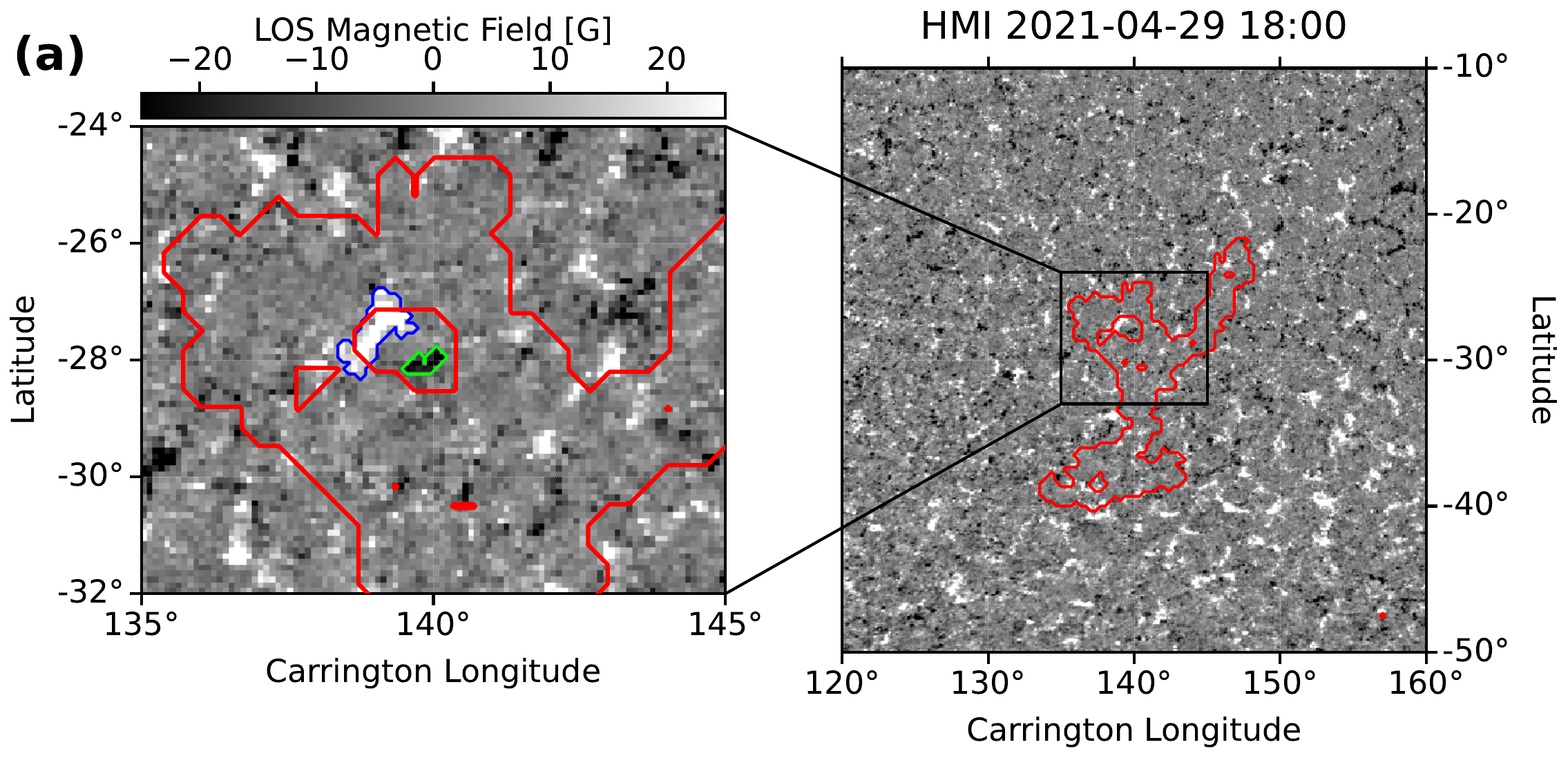} 
        %\caption{Generic} 
        %\label{fig:timing1}
    \end{subfigure}
    \hfill
    \begin{subfigure}[t]{0.495\textwidth}
        \centering
        \includegraphics[width=\linewidth]{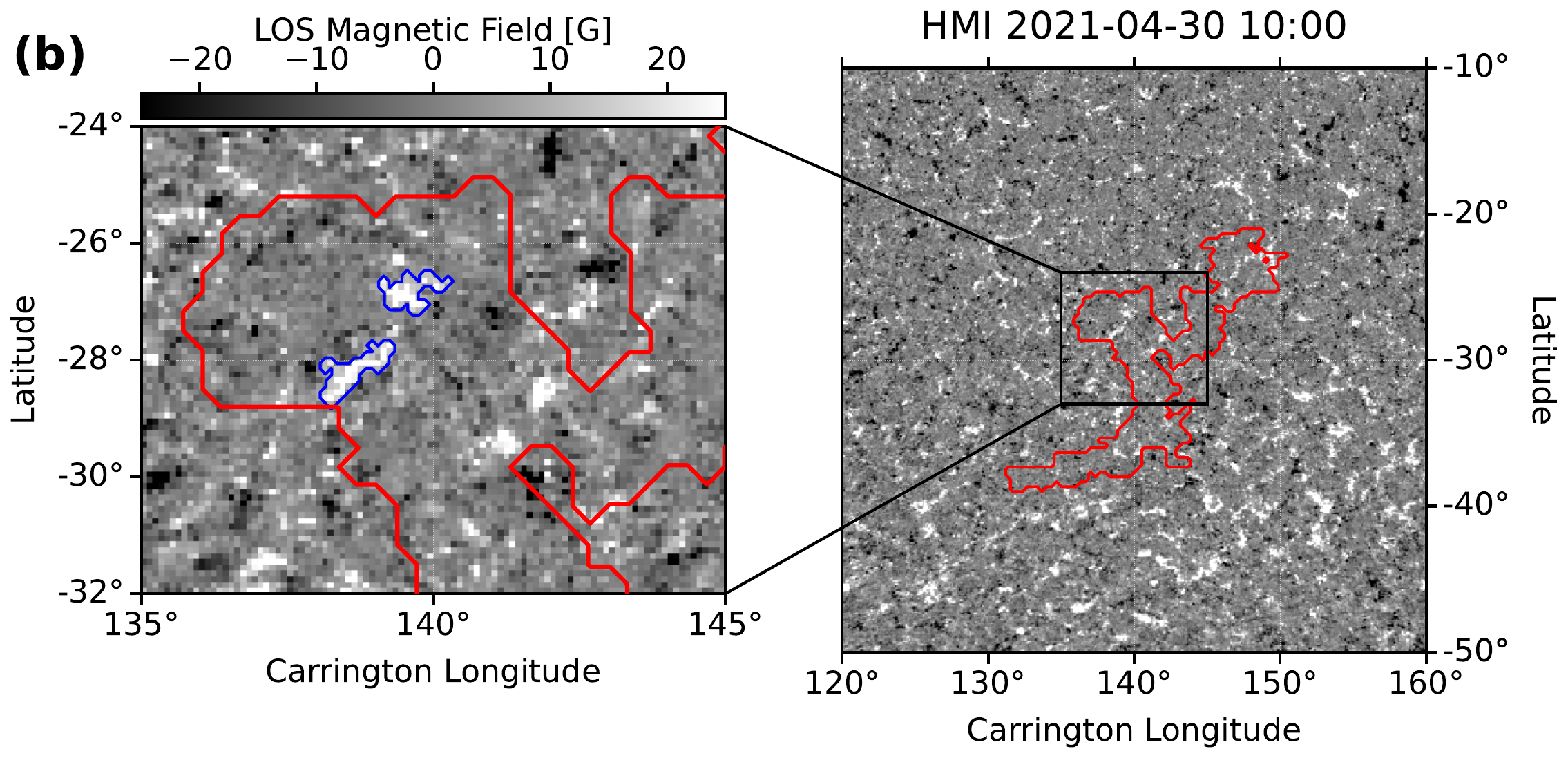} 
        %\caption{Competitors} 
        %\label{fig:timing2}
    \end{subfigure}

  %  \vspace{-1cm}
    \centering
    \begin{subfigure}[t]{0.495\textwidth}
        \centering
        \includegraphics[width=\linewidth]{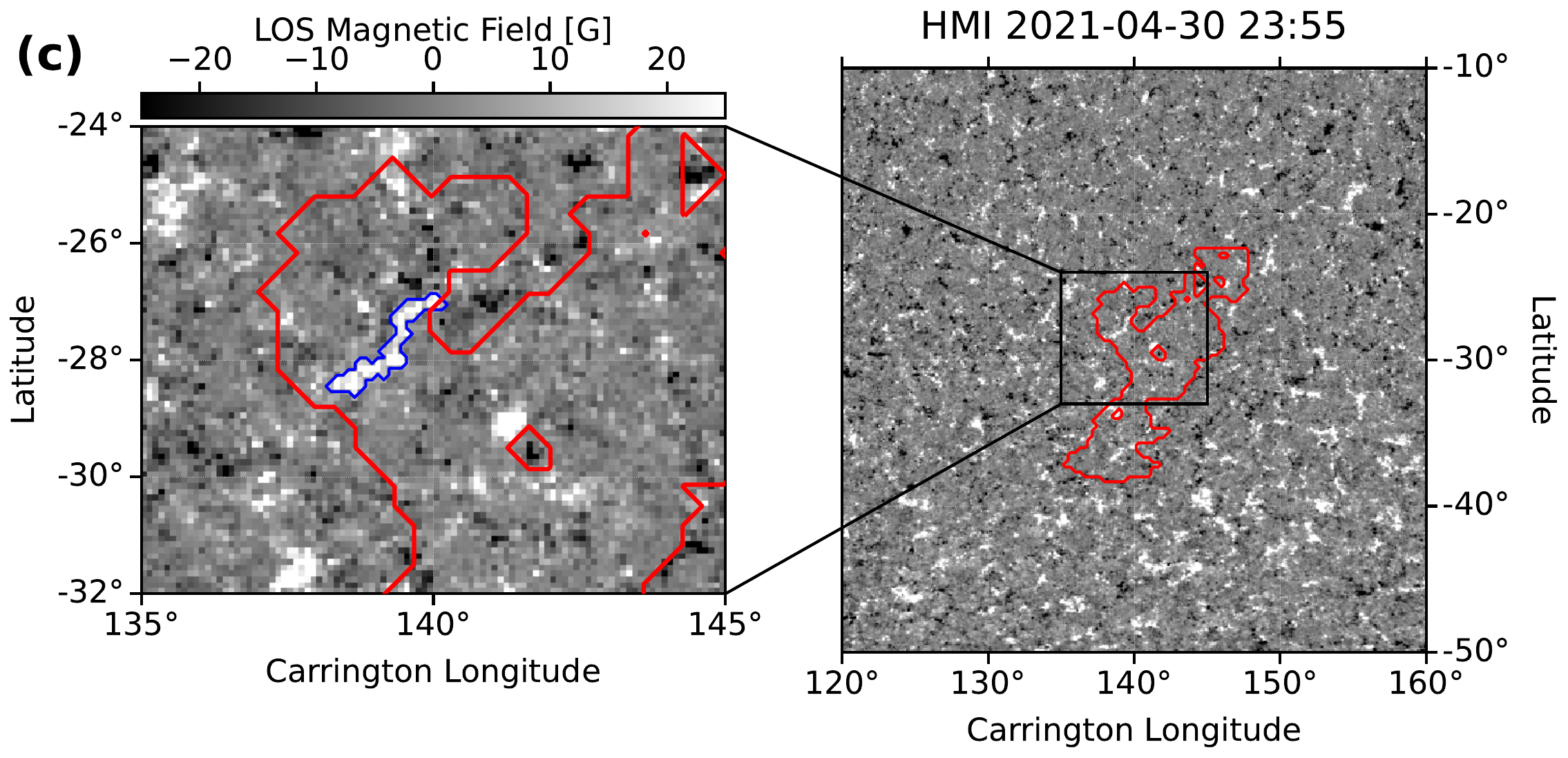} 
        %\caption{Generic} 
        %\label{fig:timing3}
    \end{subfigure}
    \hfill
    \begin{subfigure}[t]{0.495\textwidth}
        \centering
        \includegraphics[width=\linewidth]{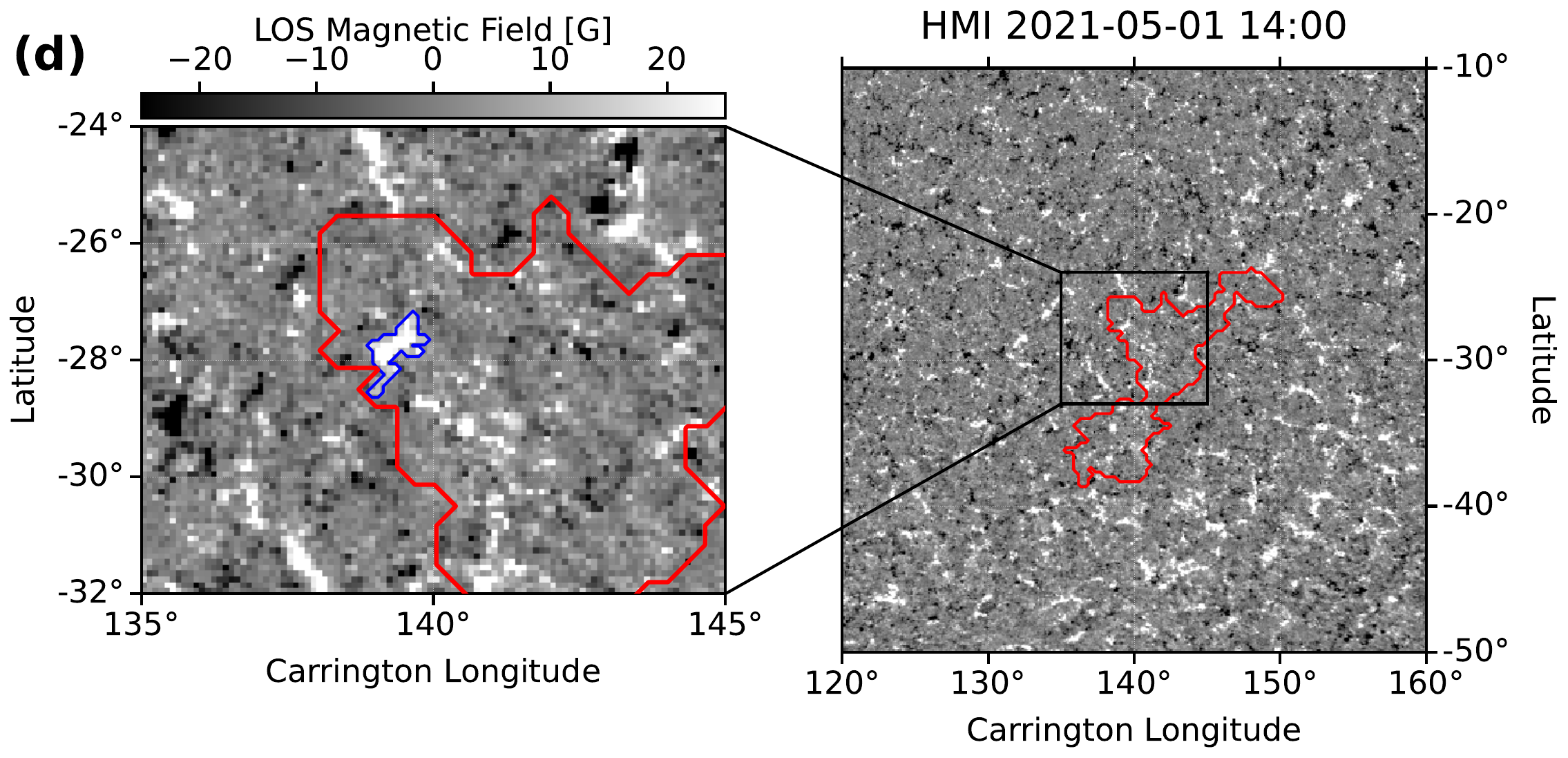} 
        %\caption{Competitors}
        %\label{fig:timing4}
    \end{subfigure}
   % \vspace{-1cm}
    \centering
    \begin{subfigure}[t]{0.495\textwidth}
        \centering
        \includegraphics[width=\linewidth]{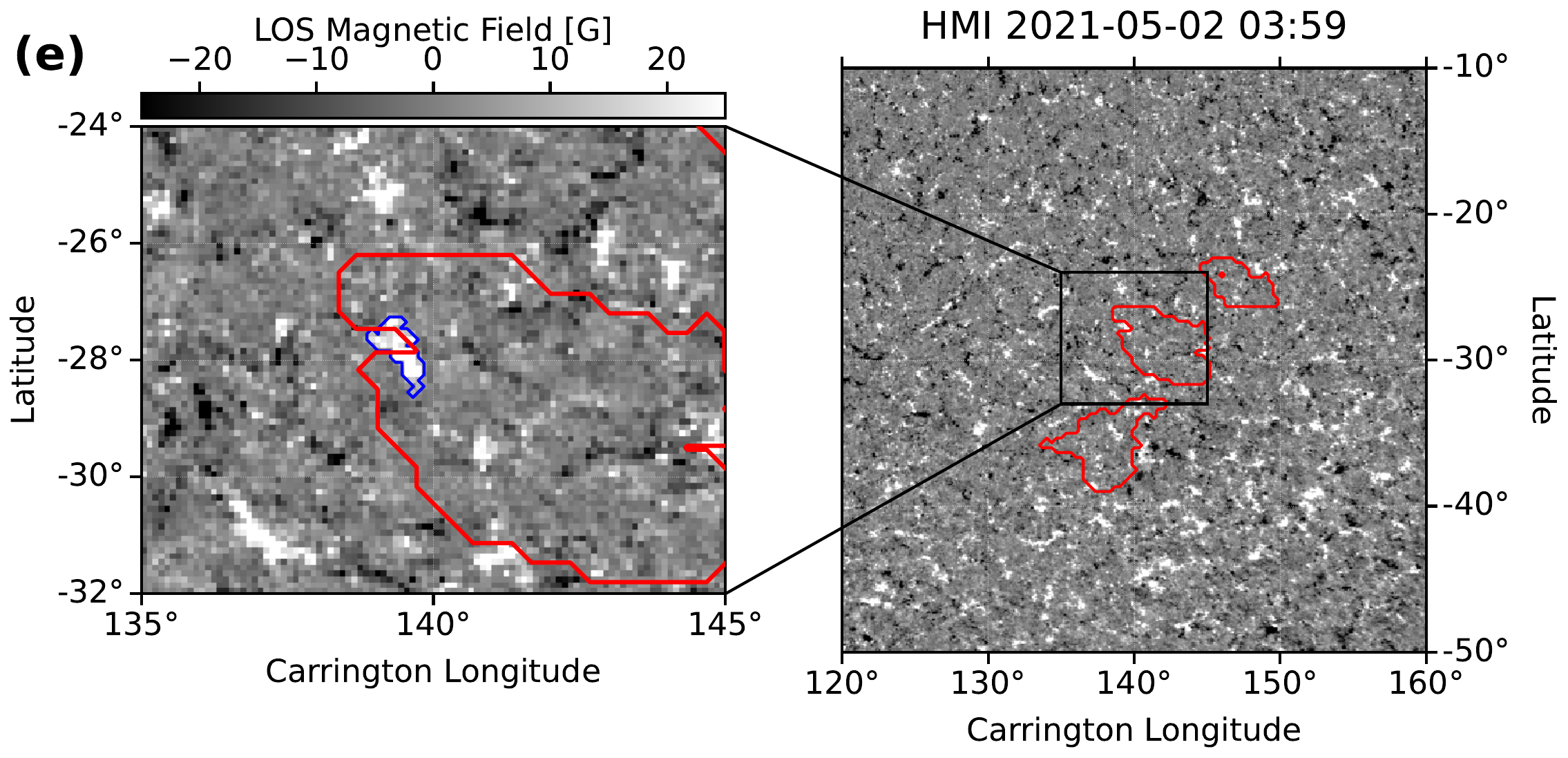} 
        %\caption{Generic} \label{fig:timing1}
    \end{subfigure}
    \hfill
    \begin{subfigure}[t]{0.495\textwidth}
        \centering
        \includegraphics[width=\linewidth]{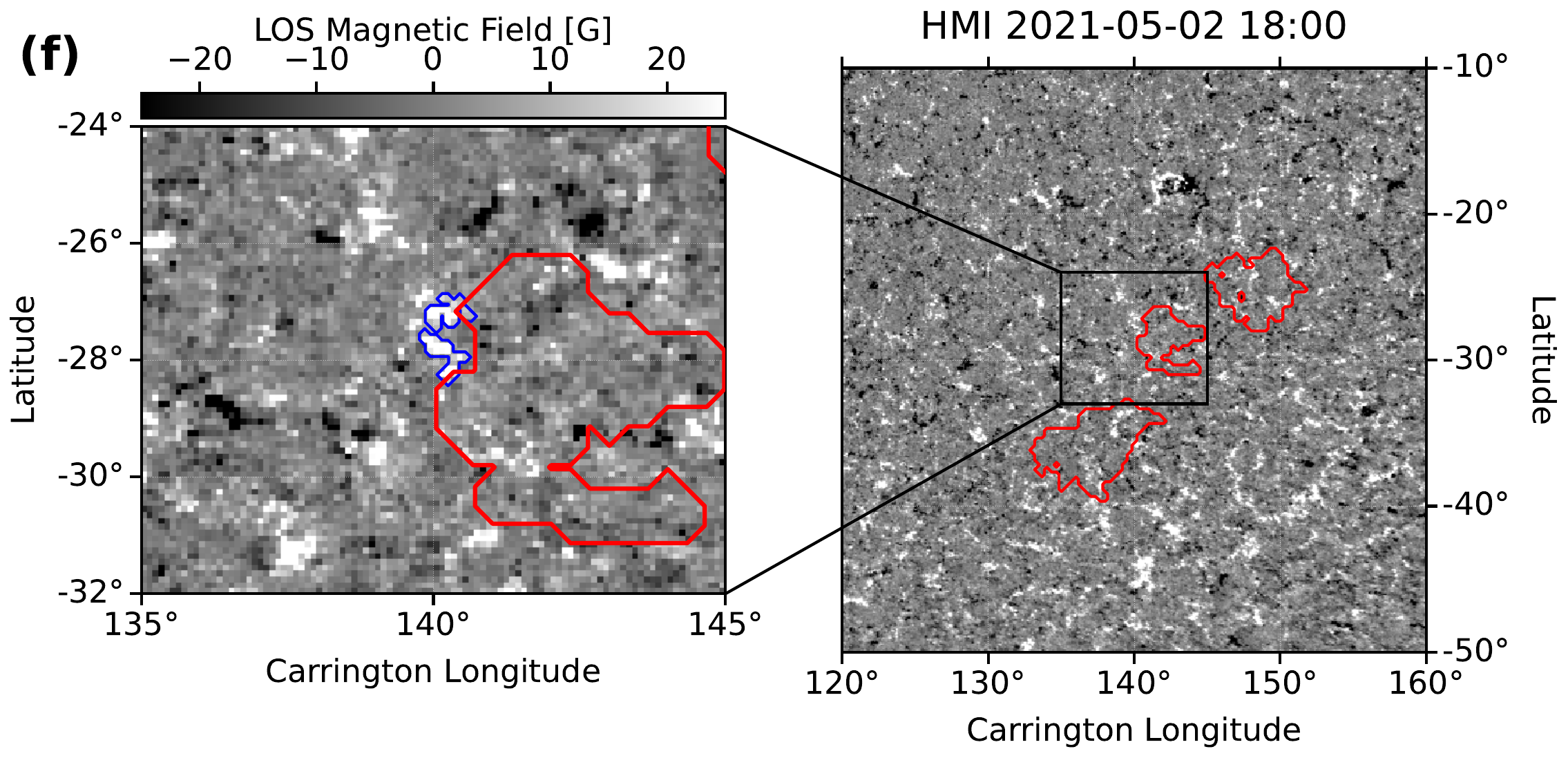} 
        %\caption{Competitors} \label{fig:timing2}
    \end{subfigure}
    
    \caption{Evolution of the LOS magnetic field underlying the coronal hole at a cadence of 14 hours. A sub-region is highlighted to show the motion of a long-living unipolar magnetic element (blue contour). The coronal hole boundary derived from EUV observations is shown in red. The green contour in panel (a) marks the location of a coronal bright point within the coronal hole (can be seen in 193 \AA ), which does not exist anymore 14 hours later.}
    \label{fig:fts}
\end{figure*}

Figure~\ref{fig:angle} shows a tilting rate of the coronal hole (up to May 1\ts{st}) that exceeds the rate expected from a coronal hole being advected by the photospheric differential rotation. This disparity between the rotation rate of the coronal hole and photosphere manifests as magnetic elements rotating out of the projected eastern (left) coronal hole boundary. In Figure~\ref{fig:fts}, we show the magnetic field evolution of a sub-region of the coronal hole over a few days. The overlaid coronal hole boundary highlights the respective area evolution in that sub-field on the coronal hole's east side. In the first timestep shown in panel (a) a small bipolar region, which corresponds to a coronal bright point in AIA 193\AA , is excluded from the extracted coronal hole boundary. This bright point cannot be observed in the next image shown, likely due to flux cancellation with the opposite polarity region next to it within less than 14 hours and only a large unipolar magnetic element remains as a footpoint of the open field and part of the coronal hole. Over the next three days this specific magnetic element is moving eastward relative to the boundary and can be seen located outside the coronal hole boundary at later times. This is consistent with a motion of the coronal hole, faster than the photospheric differential rotation. Another possible explanation is that the apparent westward motion of the coronal hole in this sub-field might be caused by the evolution (in this case closing or shrinking) of the boundary. We find that other long-living unipolar magnetic elements show an eastward motion relative to the overlying coronal hole, as well. 

\subsection{Topological evolution in magnetic field extrapolations} \label{subs:PFSS}

\begin{figure*}
   \centering
   \includegraphics[width=\linewidth]{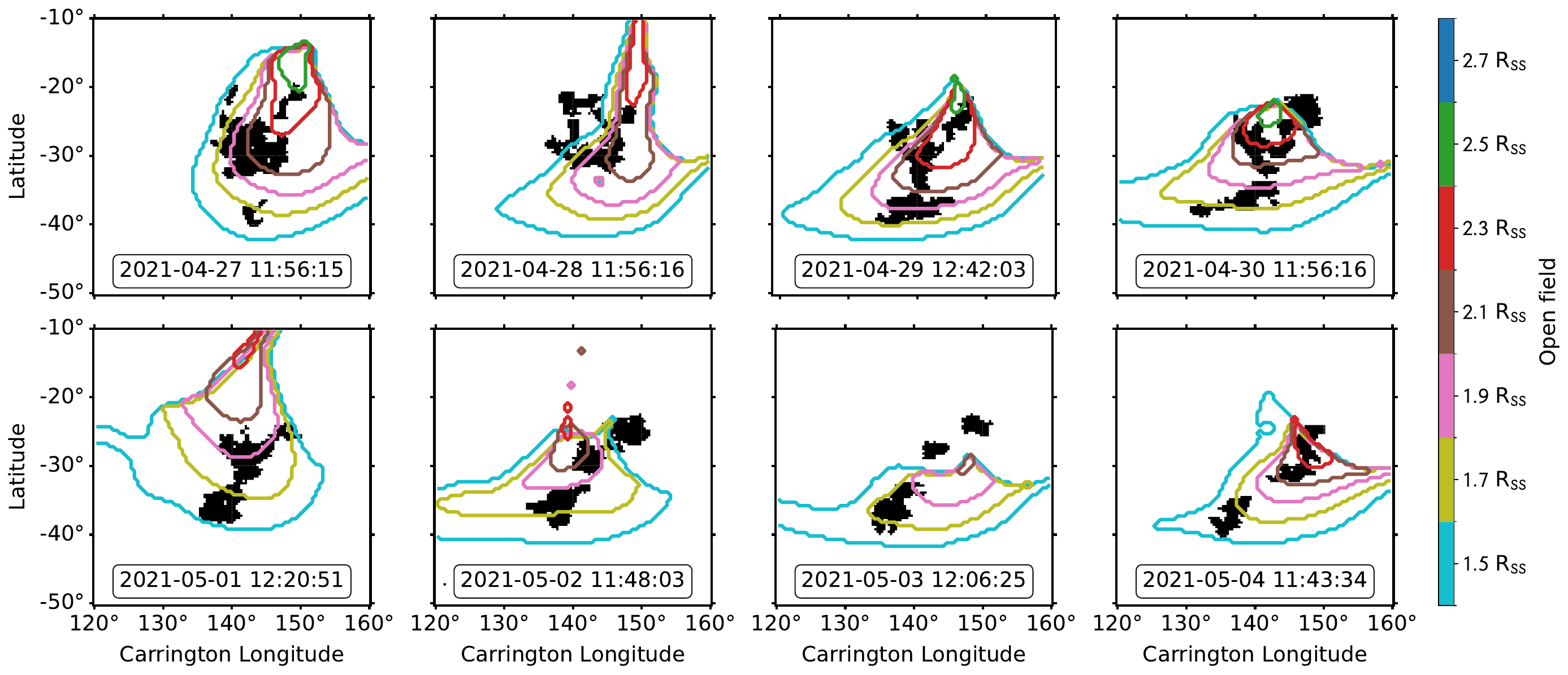}
   \caption{Coronal hole area overlayd with open field footpoints. The open fields were calculated at $1.05$ R$_{\odot}$ using PFSS modeling and different source surface heights (R$_{\mathrm{SS}}$). The colored contours represent the open field area for respective source surface heights as noted on the colorbar.}
              \label{fig:pfss}%
\end{figure*}

Coronal holes are commonly defined by their reduced emission in the corona and their open magnetic field configuration. They are often reconstructed as open field areas derived from magnetic field modeling. The most widespread approach is to consider the coronal hole approximately force free and potential. This approach uses either potential field source surface models \citep[PFSS;][]{1969altschuler_PFSS} or adaptations like the Wang-Sheeley-Arge model \citep{2000arge} that includes a \textit{Schatten} current sheet \cite{1968schatten}. To evaluate the conformity of the open fields derived from a potential magnetic field model with the observed coronal hole and to determine if the modeled open field region shows an evolutionary behavior akin to the observed structure, we model multiple instances over time using a global PFSS model. We employ a finite difference scheme \citep{Stansby2020} with synchronic HMI magnetograms (\texttt{hmi.mrdailysynframe\_polfil\_720s}).\\ 

In Figure~\ref{fig:pfss}, we present the modeled open fields calculated for different source surface heights $R_{\mathrm{ss}}$ (\textit{i.e.,} the height at which the field lines are forced to be radial). We found, that for the standard height for the source surface ($R_{ss} = 2.5\,R_{\odot}$), barely any open field areas are reconstructed. For that source surface height and from April 27th to April 30th, only a tiny patch of open field can be seen near the coronal hole's northern edge. Following this time period, no open fields are produced for this height. When reducing the value for the source surface height to around $R_{ss} = 2.1\,R_{\odot}$, we find some open fields roughly at the location of the coronal hole for all times considered. The shape and precise location of these modeled open fields, however, do not match the observed coronal hole shape. The surrounding modeled closed field structure also does not exhibit any sudden or large scale reconfiguration during this time period (not shown here). And although it is common to observe some disparity between observed coronal hole areas and modeled open fields as a function of source surface height and in terms of coronal hole shape \citep{2019asvestari,Asvestari2020,2021Linker_ISSI}, our model results lack any evidence of the evolution observed in EUV. There are no changes in the calculated open field area that resemble the changes in the observed morphology nor the observed apparent shear or tilting motion of the coronal hole presented in Figures~\ref{fig:angle} and \ref{fig:props}. Note that, the choice of magnetogram may influence the resulting open field regions \citep[\textit{e.g.,}][]{Caplan2021,Asvestari2023_ISSI2}. To account for this, we calculated the PFSS open field regions for multiple magnetograms (using GONG and ADAPT maps as well as different resolutions), but found none that matched the observations to any significantly better degree. However, this does not exclude the possibility that this discrepancy may stem from inadequate boundary conditions in the model. \\

\subsection{Dynamic evolution}
\label{ssec:chdyn}

%----------------------------------------------------------------- 
\begin{figure*}
\centering
\includegraphics[width=\linewidth]{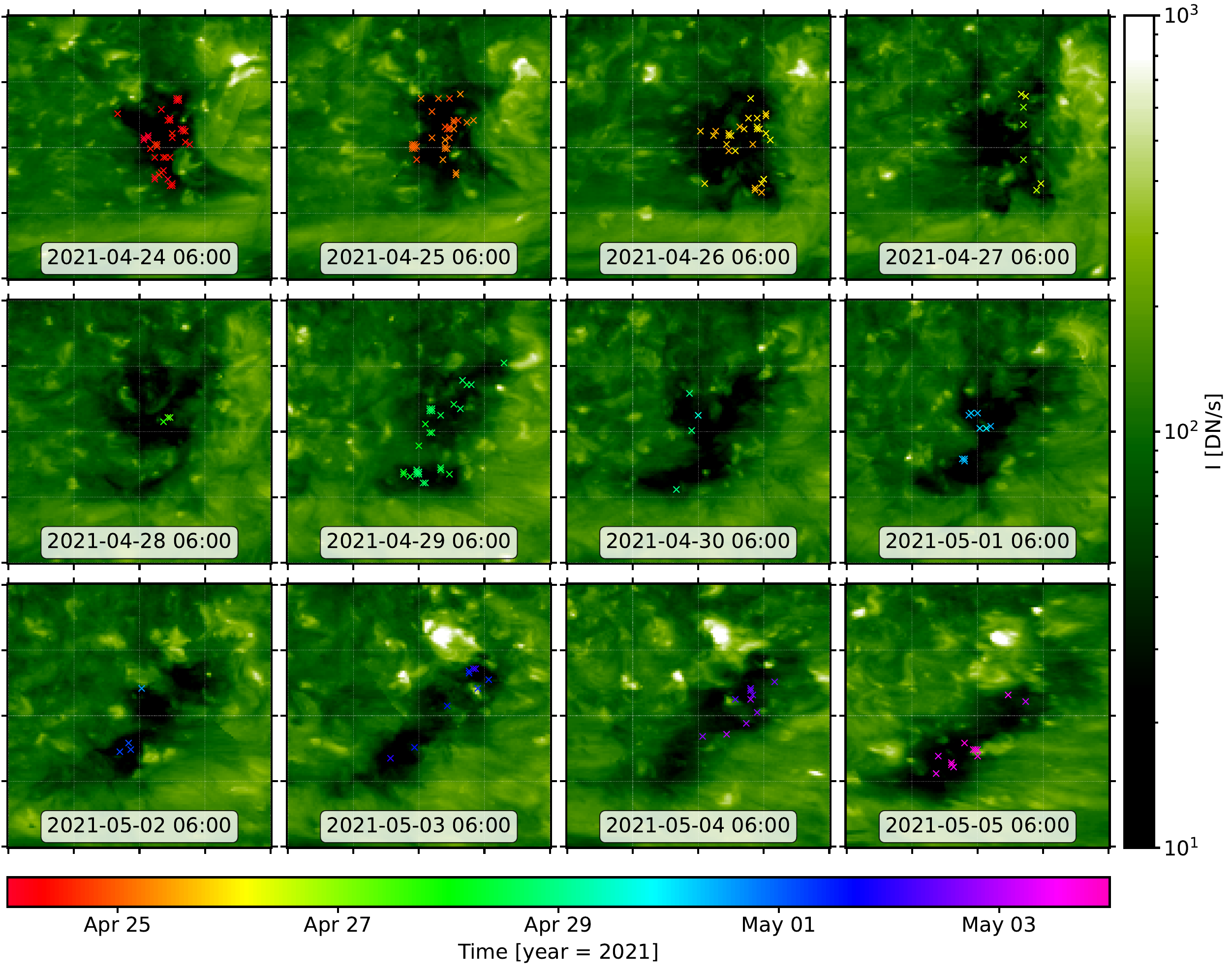}
  \caption{Daily images of the evolution of the coronal hole from April 24\ts{th}, 2021 to May 5\ts{th}, 2021. The images show merged STEREO-A/EUVI 195\AA\  and SDO/AIA 193\AA\ filtergrams as prepared by the CHMAP pipeline. The coronal hole is shown in the heliographic \textit{Carrington} frame on a $10^{\circ}$ grid (white guidelines) from $-50^{\circ}$  to $-10^{\circ}$ latitude and $120^{\circ}$  to $160^{\circ}$ longitude. In each image the locations of short-term EUV brightenings that occurred in the time period of $\pm12$ hours around the shown frame are marked as ``$x$''. The marked locations are color-coded according to the time of detection.}
     \label{fig:IR}
\end{figure*}

To investigate the dynamics of coronal hole evolution, we look for small-scale short-term EUV brightenings that are commonly regarded as tracers for reconnection events \citep[\textit{e.g.,}][]{Shokri2022}. In our dataset we define EUV brightenings as pixels where the intensity increases by more than $100\%$ in one timestep and its enhancement does not appear to last longer than 10 minutes. In addition, the intensity enhancement must be $3\sigma$ above the mean coronal hole intensity to avoid the inclusion of noise-dominated changes.\\

In Figure~\ref{fig:IR} we mark the location of all brightenings detected in and around the coronal hole color-coded as a function of time. In each panel, all tracer locations detected within $\pm 12$ hours of the times when the shown EUV image are displayed. We find that these brightenings occur continuously throughout the observed time period. Preferred locations are where changes in the coronal hole are seen in subsequent images. Around May 3\ts{rd} to May 5\ts{th}, we find enhancements in the northern part of the coronal hole that is in close proximity to the emerging ephemeral region. We detected 205 such brightenings in the observed time period. Between April 24\ts{th} and 26\ts{th}, we find an occurrence rate of just above $1$ per hour, while afterwards the rate drops to 1 per 2-3 hours. We suspect that the initial higher rate (during times that are dominated by STEREO-A observerations) of brightenings might be due to some instrumental differences between STEREO and SDO that remains even after the inter-instrument calibration. \\

\section{Discussion}
\label{sec:disc}
In Section~\ref{sec:obs} we presented the evolution of a coronal hole over 12 days. Until May 1\ts{st}, the coronal hole shows a clockwise tilting motion that appears to develop at a rate slightly higher than the expected value if the sole source of the tilting was advection of its phtospheric footpoints according to a differential rotation profile. Around May 2\ts{nd} to May 4\ts{th} the coronal hole starts tilting rapidly at nearly double the previous rate. In addition, during the same time period from April 30\ts{th} to May 3\ts{rd}, the area decreases by more than a factor of three within a few days while at the same time the majority of the coronal hole's open flux is conserved.\\

It is believed that the open field that is seen as dark regions in the solar corona is rooted in unipolar magnetic elements in the photosphere \citep{2019hofmeister}. According to the general understanding of the expansion of these open fields \citep[][]{2009wedemeyer,2019cranmer}, it can be assumed that their coronal representation (\textit{i.e.} coronal hole) is formed by the sum of the open field lines extending from their footpoints and the interaction with the surrounding and global fields. It is well known that the photosphere and as such the magnetic elements on large scales rotate differentially, which would suggest that a coronal hole should rotate in the same manner as those magnetic elements if an inflexible connection between footpoints and the corona is assumed. However, this  contradicts the observed rigid behavior \citep[\textit{e.g.,}][]{1975timothy} of coronal holes not rotating entirely differentially. \cite{2018heinemann_paperI} found a tilting rate for one specific coronal hole in 2012 of $0.42^{\circ}\mathrm{day}^{-1}$ which is at least a factor of four smaller than the photospheric rotation rate for the same latitudinal range due to differential rotation \citep[as given by][]{Snodgrass1990}, and around half of the rotation rate of coronal holes as derived by \cite{2017bagashvili}. A rotation rate slower than differential rotation could be caused by the faster rotating underlying photosphere applying a twist on the more uniform magnetic coronal structure causing footpoint switching near the boundary to retain the coronal hole's integrity \citep[\textit{e.g.,}][]{2018kong}.\\

In our study, we show that (1) the tilting rate appears to exceed the rate predicted by a coronal hole solely advected by a photospheric differential rotation profile. The excess tilting is even larger when comparing to a coronal rotation profile (see Fig.~\ref{fig:angle}). And (2) the suspected coronal hole footpoints (unipolar magnetic elements) move eastward in comparison to the coronal hole boundary, which indicates that parts of the coronal hole (as seen in the corona) rotate faster than the photospheric differential rotation (see Fig.~\ref{fig:fts}). And lastly (3), we find that the PFSS-modeled open-field structure does not show any evolutionary behavior akin to the structure observed in EUV. The noticeable tilting motion can not be seen in the modeled open fields (see Fig.~\ref{fig:pfss}). From these three observations, it follows that the rate at which the coronal hole tilts cannot be explained by photospheric and/or coronal differential rotation acting on the coronal hole. Thus it becomes necessary to identify alternative physical mechanisms responsible for this phenomenon. We propose two possibilities:

Firstly, we cannot discern whether the strong apparent tilting motion, starting around May 2\ts{nd}, was caused by coronal hole substructures evolving individually. The evolution of the northernmost parts of the coronal hole seems to be strongly influenced by the emergence of a nearby ephemeral active region. The sudden flux emergence near the boundary may have led to a westward shifting of a part of the coronal hole. This can occur due to reconnection to an energetically lower state or due to being pushed by magnetic pressure. \cite{Terradas2023} showed such a case using 2D magnetostatic simulations. \\

Secondly, we suspect that continuous interchange reconnection in and around the coronal hole \citep[][]{2009madjarska} caused dynamic changes in the open field configuration which could account for the observed apparent tilt motion. Indicators for this are the magnetic elements that are seen moving in and out of the coronal hole in an eastward direction and that the open flux in the coronal hole remains roughly constant until after May 3\ts{rd} although the area has decreased by approximately a factor of three. This suggests that the majority of the morphological evolution and area decay is caused by a flux conserving process, likely interchange reconnection in the corona, rather than changes in the underlying photospheric magnetic field structure due to flux emergence and cancellation. We support this, by showing that localized short-term EUV brightenings can be seen during the coronal holes evolution (see Fig.~\ref{fig:IR}). Those brightenings are primarily located around areas that show subsequent morphological changes. They may indicate coronal transient events observed at these spatial scales such as coronal jets \citep{Sterling2015,Panesar2018}, coronal bright points \citep{Habbal1981,Matkovi2023}, or blinkers \citep{Subramanian2012,Shokri2022}; furthermore it is generally acknowledged that reconnection plays a major role in their formation \citep[\textit{e.g.,}][]{Shokri2022}. 
Additionally, a coronal hole features a primarily open-close magnetic configuration with lower lying closed loops and open fields stretching out into the heliosphere \citep[\textit{e.g.,}][]{2004wiegelmann,bale2023}. Thus, it is reasonable to speculate that these detected brightenings are connected to interchange reconnection with associated small-scale restructuring of the open-closed field.\\

\section{Conclusions}
\label{sec:sum}

In this study we presented the evolution of a well-observed coronal hole over 12 consecutive days. Using EUV and LOS magnetic field observations in combination with PFSS modeling, we tracked and analyzed the morphological and magnetic evolution of the coronal hole from April 23\ts{rd} to May 5\ts{th} 2021. Our findings can be summarized as follows:

\begin{itemize}
    \item For the first time, we describe a coronal hole that tilts at a rate exceeding the rate expected from differential rotation. Our coronal hole tilts at a mean rate ($3.2^{\circ}$day$^{-1}$) that exceeds the expected effect of solar plasma being advected by either photospheric ($2.6^{\circ}$day$^{-1}$) or coronal hole differential rotation ($1.2^{\circ}$day$^{-1}$).
    
    \item In the time period from May 1\ts{st} to May 3\ts{rd}, we observed an acceleration of the tilting rate to $5.4^{\circ}$day$^{-1}$.
    
    \item While the area of the coronal hole decays by more than a factor of three in approximately four days (from $\sim~12.8 \times 10^{9}$\,km$^{2}$ to $\sim~3.9 \times 10^{9}$\,km$^{2}$), its signed flux remains conserved at a value of $(1.82 \pm 0.43) \times 10^{20}$\,Mx. 

    \item We find that photospheric unipolar magnetic elements move eastward relative to the overlying coronal hole boundary. This hints that parts of the coronal hole rotate (or evolve) faster than the underlying photosphere.
    
    \item The prominent tilting motion of the coronal hole can not be reproduced by potential field extrapolations of the coronal hole's open field structure.

    \item Throughout the observed time period, we find small-scale brightenings occurring in and around the coronal hole at a rate of $0.3$ to $1$ per hour. These are suspected to be tracers for interchange reconnection.

    \item We suggest that the rapid morphological evolution of the coronal hole may be primarily driven by interchange reconnection at the boundary (which is a flux conserving process) and the interaction with a newly emerging ephemeral region in the northern part of the coronal hole. 
    
\end{itemize}

This coronal hole highlights the dynamic nature of the solar corona even in regions traditionally believed to be rigid and stable. The dynamic evolution may have a significant effect on the outflowing solar wind and could also be a reason for the poor agreement between EUV observations and PFSS modeled open fields.

\begin{acknowledgements}
This paper resulted from discussions with the team of the extensive multi-instrument observation campaign lead by S.~J.~Hofmeister from  April 28th 2021. Thus, the authors would like to thank P.~Bryans, M.~Cantoresi, D.~J.~Christian, M.~Globa, M.~Hahn, N.~Huang, C.~Kuckein, D.~Lacatus, G.~de~Toma, K.~Tziotziou, E.~Samara, C.~Vocks and D.~W.~Savin for fruitful discussions and helpful comments. The authors would like to thank J.~Linker and R.~Caplan from Predictive Science Inc. for providing the coronal hole extractions. SGH expresses gratitude to K.~Horaites for engaging in inspirational listening. This research was funded in whole, or in part, by the Austrian Science Fund (FWF) Erwin-Schr\"odinger fellowship J-4560. For the purpose of open access, the author has applied a CC BY public copyright licence to any Author Accepted Manuscript version arising from this submission. SJH acknowledges support by the German Science Fund (DFG), grant No. 448336908. ACS was supported with funding from the Heliophysics Division of NASA’s Science Mission Directorate through the Heliophysics Supporting Research (HSR, grant No. 20-HSR20 2-0124) Program, and through the Heliophysics System Observatory Connect (HSOC, grant No. 80NSSC20K1285) Program. JAT and CD acknowledge support in the creation of CHMAP from the NASA Guest Investigators Program (NNX17AB78G). JP acknowledges the Academy of Finland Project 343581. EA acknowledges support from the Academy of Finland (Postdoctoral Researcher Grant 322455) and the ERC under the European Union's Horizon 2020 Research and Innovation Programme Project 724391 (SolMAG). The authors acknowlege support from NASA Heliophysics Living with a Star grant 80NSSC20K0183. The SDO and STEREO image data are available by courtesy of NASA and the respective science teams.  
\end{acknowledgements}

\bibliographystyle{aa} % style aa.bst
  % \bibliography{bib} % your references Yourfile.bib

%-------------------------------------------------------------------
\end{document}